\documentclass[twoside,twocolumn,9pt]{article}
\usepackage{extsizes}
\usepackage{bm}
\usepackage[super,sort&compress,comma]{natbib} 
\usepackage[version=3]{mhchem}
\usepackage[left=1.5cm, right=1.5cm, top=1.785cm, bottom=2.0cm]{geometry}
\usepackage{balance}
\usepackage{mathptmx}
\usepackage{sectsty}
\usepackage{graphicx} 
\usepackage{lastpage}
\usepackage[format=plain,justification=justified,singlelinecheck=false,font={stretch=1.125,small,sf},labelfont=bf,labelsep=space]{caption}
\usepackage{accents}

\usepackage{float}
\usepackage{fancyhdr}
\usepackage{fnpos}
\usepackage[english]{babel}
\addto{\captionsenglish}{%
  
}
\usepackage{array}
\usepackage{droidsans}
\usepackage{charter}
\usepackage[T1]{fontenc}
\usepackage[usenames,dvipsnames]{xcolor}
\usepackage{setspace}
\usepackage[compact]{titlesec}
\usepackage{hyperref}
\usepackage{amsmath,amssymb}
\DeclareMathOperator{\sign}{sgn}

\definecolor{cream}{RGB}{222,217,201}

\begin{document}

\pagestyle{fancy}
\thispagestyle{plain}
\fancypagestyle{plain}{
\renewcommand{\headrulewidth}{0pt}
}

\makeFNbottom
\makeatletter
\renewcommand\LARGE{\@setfontsize\LARGE{15pt}{17}}
\renewcommand\Large{\@setfontsize\Large{12pt}{14}}
\renewcommand\large{\@setfontsize\large{10pt}{12}}
\renewcommand\footnotesize{\@setfontsize\footnotesize{7pt}{10}}
\makeatother

\renewcommand{\thefootnote}{\fnsymbol{footnote}}
\renewcommand\footnoterule{\vspace*{1pt}%
\color{cream}\hrule width 3.5in height 0.4pt \color{black}\vspace*{5pt}} 
\setcounter{secnumdepth}{5}

\makeatletter 
\renewcommand\@biblabel[1]{#1}            
\renewcommand\@makefntext[1]%
{\noindent\makebox[0pt][r]{\@thefnmark\,}#1}
\makeatother 
\renewcommand{\figurename}{\small{Fig.}~}
\sectionfont{\sffamily\Large}
\subsectionfont{\normalsize}
\subsubsectionfont{\bf}
\setstretch{1.125} 
\setlength{\skip\footins}{0.8cm}
\setlength{\footnotesep}{0.25cm}
\setlength{\jot}{10pt}
\titlespacing*{\section}{0pt}{4pt}{4pt}
\titlespacing*{\subsection}{0pt}{15pt}{1pt}

\fancyfoot{}
\fancyfoot[LO,RE]{\vspace{-7.1pt}\includegraphics[height=9pt]{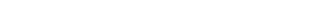}}
\fancyfoot[CO]{\vspace{-7.1pt}\hspace{13.2cm}\includegraphics{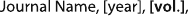}}
\fancyfoot[CE]{\vspace{-7.2pt}\hspace{-14.2cm}\includegraphics{head_foot/RF}}
\fancyfoot[RO]{\footnotesize{\sffamily{1--\pageref{LastPage} ~\textbar  \hspace{2pt}\thepage}}}
\fancyfoot[LE]{\footnotesize{\sffamily{\thepage~\textbar\hspace{3.45cm} 1--\pageref{LastPage}}}}
\fancyhead{}
\renewcommand{\headrulewidth}{0pt} 
\renewcommand{\footrulewidth}{0pt}
\setlength{\arrayrulewidth}{1pt}
\setlength{\columnsep}{6.5mm}
\setlength\bibsep{1pt}

\makeatletter 
\newlength{\figrulesep} 
\setlength{\figrulesep}{0.5\textfloatsep} 

\newcommand{\topfigrule}{\vspace*{-1pt}%
\noindent{\color{cream}\rule[-\figrulesep]{\columnwidth}{1.5pt}} }

\newcommand{\botfigrule}{\vspace*{-2pt}%
\noindent{\color{cream}\rule[\figrulesep]{\columnwidth}{1.5pt}} }

\newcommand{\dblfigrule}{\vspace*{-1pt}%
\noindent{\color{cream}\rule[-\figrulesep]{\textwidth}{1.5pt}} }

\makeatother

\twocolumn[
  \begin{@twocolumnfalse}
{\includegraphics[height=30pt]{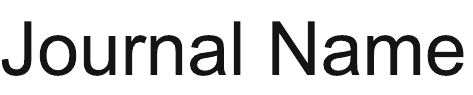}\hfill\raisebox{0pt}[0pt][0pt]{\includegraphics[height=55pt]{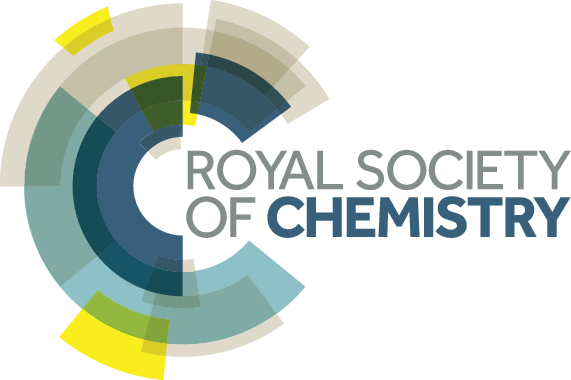}}\\[1ex]
\includegraphics[width=18.5cm]{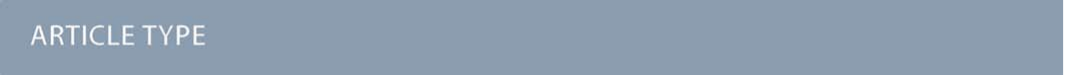}}\par
\vspace{1em}
\sffamily
\begin{tabular}{m{4.5cm} p{13.5cm} }

\includegraphics{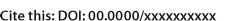} & \noindent\LARGE{\textbf{Soft wetting ridge rotation in sessile droplets and capillary bridges}} \\
\vspace{0.3cm} & \vspace{0.3cm} \\

 & \noindent\large{Bo Xue Zheng$^{a}$ and Tak Shing Chan$^{a*}$} \\

\includegraphics{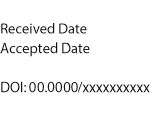} & \noindent\normalsize{
 We study the deformation of soft solid layers  in the presence of sessile droplets or  capillary bridges. By incorporating the  surface tension balance at the contact line, we examine the rotation of the wetting ridge and the corresponding change in the contact angle. Our findings reveal that the rotation direction of the wetting ridge aligns with the sign of the Laplace pressure.  Notably, while a softer solid layer typically decreases the contact angle for sessile droplets, a negative Laplace pressure in a hydrophilic capillary bridge pulls the solid-liquid interface, leading to an increased contact angle. The interplay between soft layer deformation and droplet contact angle modulation offers  insights for controlling droplet motion through elastocapillarity.} \\  


\end{tabular}

 \end{@twocolumnfalse} \vspace{0.6cm}

  ]

\renewcommand*\rmdefault{bch}\normalfont\upshape
\rmfamily
\section*{}
\vspace{-1cm}


\footnotetext{\textit{$^{a}$~Mechanics Division, Department of Mathematics, University of Oslo, 0316 Oslo, Norway. Email: taksc@uio.no }}


\section{Introduction}

Manipulating surface wettability to control droplet morphology and movement \cite{Tenjimbayashi2022} has broad applications in  industrial processes and is widely observed in natural phenomena. Over the past decades, extensive research has focused on droplets in contact with substrates coated with soft material layers \citep{Pericet-Camara2008,Jerison2011,Limat2012,Yu2012,Style2012,Style2013prl,Kajiya2013,Park2014,Bostwick2014,Karpitschka2015,Style2017,Fernandez-Toledano2017,Andreotti2020,Pandey2020,Chan2022,Zheng2023} such as gels and elastomers. Studies have shown that manipulating the softness of the layer can provide an alternative way to control the motion of droplets \cite{Style2013pnas,Liu2017,Bueno2018,Bardall2020,Gomez2020,Henkel2021,Tamim2021,Kajouri2024}. Style et al.  \cite{Style2013pnas} demonstrated experimentally that a sessile droplet moves spontaneously from thinner to thicker regions of a soft layer. This phenomenon, known as durotaxis,  originally describes the process that biological cells move along gradients in stiffness of a substrate. The movement of the droplet is explained in terms of a decrease in the contact angle with the thickness of the elastic layer. The thickness gradient creates a variation in the contact angle along the contact line which drives the motion.  

  \begin{figure}
\begin{center}
\includegraphics[width=0.4\textwidth]{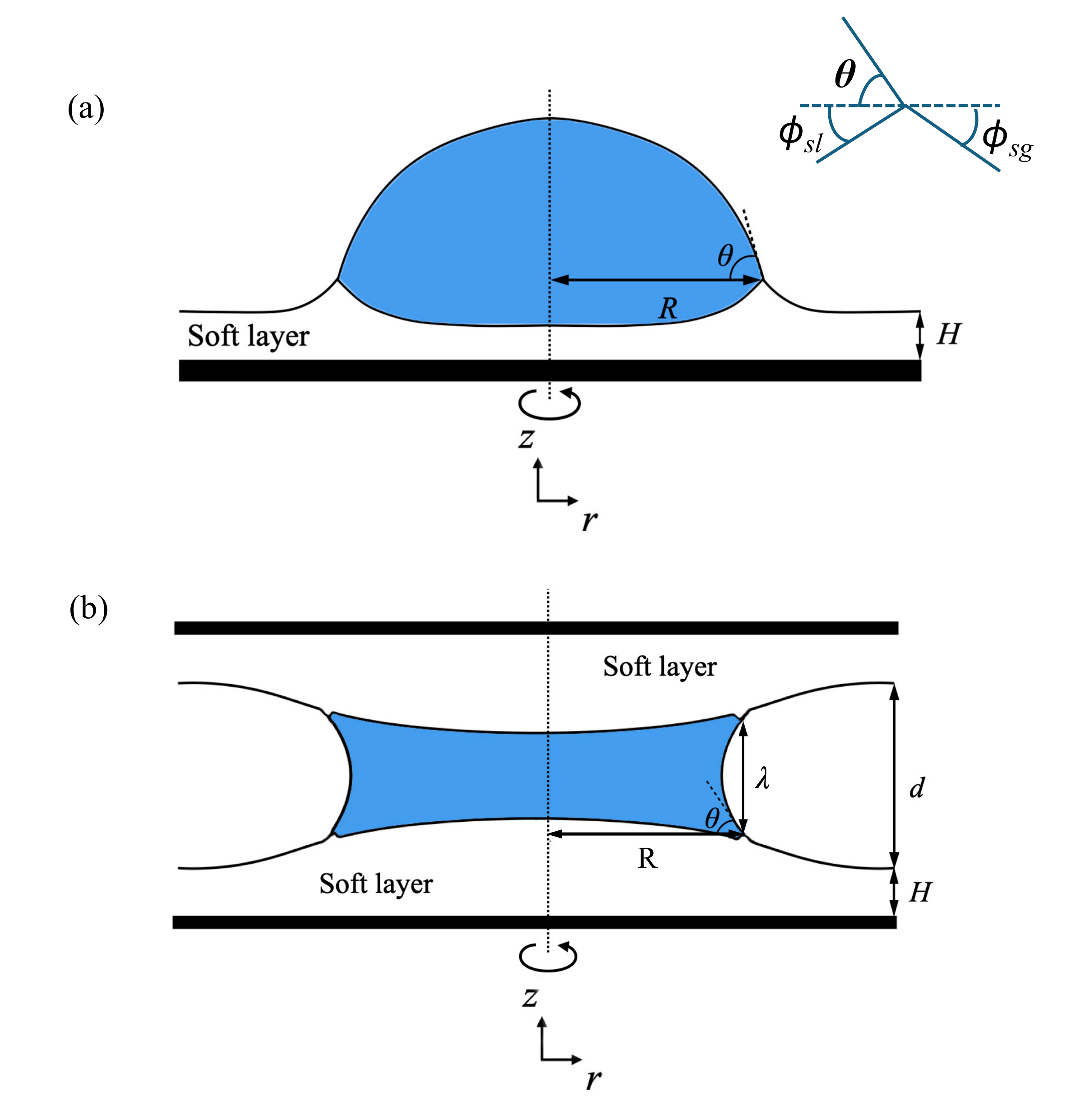}
\caption{Schematic diagrams of an axisymmetric droplet in contact with one single substrate in (a), i.e. a sessile droplet; and two parallel substrates in (b), i.e. a capillary bridge. The substrates consist of rigid plates coated with a soft elastic  layer of thickness $H$. The soft layer is deformed by the droplet to form a wetting ridge. At the contact line position, the sessile droplet or the capillary bridge makes an equilibrium contact angle $\theta$ with the substrates. The angles of the interfaces, $\theta$, $\phi_{sl}$ and $\phi_{sg}$, follow the balance of surface tensions, also known as Neumann's triangle. } \label{fig1}
\end{center}
\end{figure}
 
  For droplets in contact with a rigid substrate, the contact angle $\theta$ is often assumed to follow Young's law: i.e. $\theta=\theta_Y$, with the Young's angle $\theta_Y$ determined by 
\begin{eqnarray} \label{Ylaw}
\gamma\cos\theta_Y=\gamma_{sg}-\gamma_{sl},
\end{eqnarray}
where $\gamma$, $\gamma_{sg}$ and $\gamma_{sl}$ are respectively the liquid-gas, solid-gas and solid-liquid surface tensions. 
   However, when a droplet contacts a soft substrate,  see Fig. \ref{fig1}, the soft layer deforms to form a ridge-like shape due to a pulling capillary force at the contact lines  \citep{Yu2009,Style2017,Andreotti2020}. Minimizing the free energy \citep{Snoeijer2018}, which includes both the surface energies and the elastic energy, reveals that the angles between the interfaces at the contact line follow the balance of the surface tensions \citep{Snoeijer2018,Style2013prl}, also known as Neumann's triangle. This condition leaves one degree of freedom for determining the angles, as the surface tension vectors can be rotated by the same angle without violating the force balance. As a result, the angles are fixed only when considered together with deformation away from the contact line.

   Studies of sessile droplet with $\theta_Y=90^{\circ}$ \citep{Lubbers2014,Dervaux2015} show that the contact angle decreases significantly from the Young's angle when increasing the softness parameter $S$ for $S\gtrsim 1$, where $S\equiv \gamma_s /ER$, is defined as the ratio of two length scales: the elastocapillary length $\gamma_s/E$ and  the contact radius of the droplet $R$. Here $\gamma_s$ is the solid surface tension, assumed to be the same on the solid-liquid and the solid-gas sides, and $E$ is the Young's modulus. In the limit of very soft case, i.e. $S\gg 1$, droplets appear as a lens floating on a liquid bath for which surface tensions completely dominate over elastic stresses \citep{Lubbers2014,Dervaux2015}.

 Despite remarkable findings, most studies on soft wetting are restricted to sessile droplets and $\theta_Y=90^{\circ}$. It remains unclear how the contact angle is modified in other geometrical confinements. Even on planar surfaces, a droplet in contact with two parallel plates forms a capillary bridge \cite{FORTES1982,CARTER1988,Marmur1993,Herminghaus2005,Rabinovich2005,Qian2006,DeSouza2008,Butt2009,Chen2013}, which has a significantly different shape from a sessile droplet. Studies of capillary bridge between two soft layers are rather limited. It has been shown that the soft layer is drawn towards the bridge by the pulling contact line force and the negative Laplace pressure \cite{Wexler2014, Li2014}. In those studies, the authors either consider only the completely wetting case \cite{Wexler2014}, i.e. $\theta_Y=0^{\circ}$, or assume a smooth solid interface at the contact line position and a local Young's law \cite{Li2014}. 

 In this article, we impose the condition of surface tension balance at the contact line and examine both hydrophilic and hydrophobic surfaces. We unravel the morphology of the wetting ridge for droplets in contact with planar soft layers in two common cases: sessile droplets and capillary bridges. Our results demonstrate how the contact angle depends on the geometric parameters and the softness of the solid layers.

 


 \section{Formulation}
 We consider substrates that consist of a rigid plate coated with a soft elastic layer of uniform thickness $H$ at an undeformed state. A droplet of volume $V$ and density $\rho$ is placed in contact with 1) one single substrate to form a sessile droplet, and 2) two parallel substrates to form  a capillary bridge, as shown in the schematics in  Fig. \ref{fig1} (a) and (b) respectively. The schematics also illustrate the deformation of the soft layers due to the droplet/bridge Laplace pressure and the pulling capillary force at the contact lines. For the bridge case, the gap separation between the two undeformed soft layers is $d$, which is assumed to be much smaller than the radius of the contact line $R$, i.e. $d\ll R$. The distance between the contact lines at  the top and at the bottom is denoted as $\lambda$.  Due to axisymmetry of the problem,  we will use the cylindrical coordinate system $(r,z, \phi)$.  The droplet/bridge makes a contact angle $\theta$ with the soft layers at the contact line positions $r=R$.  Note that the contact angle is measured with respect to the plane parallel to the undeformed soft layers. The contact angle is an unknown variable that has to be determined together with the solution of the solid interface deformation. Instead of using the Young's law (eq. \ref{Ylaw}), we use the condition of balance of surface tensions at the contact line which will be described  in details at the later part of this section. We denote the displacement in the soft layers as $\bm{U}(r,z)=U_r\hat{\bm{r}}+ U_z\hat{\bm{z}}+U_{\phi}\hat{\bm{\phi}}$. Employing  linear elasticity, the relation between  the stress tensor $\bm{\sigma}$ and the strain tensor $\boldsymbol{\epsilon}=\left[\nabla\bf{U}+(\nabla\bf{U})^T \right]/2$ is given by  
 \begin{eqnarray} 
\boldsymbol{\sigma}=-p\mathbf{I}+\frac{E}{1+\nu}\left[\boldsymbol{\epsilon} - \frac{\mathrm{Tr}( \boldsymbol{\epsilon})\mathbf{I}}{3}\right],
\end{eqnarray} 
where $p$ is the pressure in the elastic layer, $\nu$ is the Poisson ratio, $\mathbf{I}$ is the identity tensor and  $\mathrm{Tr}$ represents taking the trace of a tensor.  In Appendix \ref{appA}, we give the relations between the components of the tensors in cylindrical coordinates. Denoting the gravitational acceleration as $g$, we consider that the Bond number $Bo\equiv \rho gV^{2/3}/\gamma\ll 1$,  and thus the effect of gravity on the droplet/bridge shape is negligible.  We consider static states, the deformation in the soft layer is governed by the force balance equation: 
\begin{eqnarray} \label{elst5}
\bigtriangledown\cdot \boldsymbol{\sigma}  =0.
\end{eqnarray}

For the bridge case, the deformations of the top and the bottom soft layers are the same and hence we only focus on the bottom one. We consider a rectangular domain of computation with $r=[0,L]$ and $z=[0, H]$, which represents the undeformed shape (reference state) of the soft layer.   The boundary conditions are as follows.  At a distance $r=L$ far away from the droplet/bridge, we impose the condition $\bm{U}(r=L, z)=0$.  At $r=0$,  we have $U_r=0$ and $\partial U_z/\partial r=0$ due to symmetry.  At the interface where the soft layer is in contact with the rigid substrate, i.e. $z=0$, the soft material is undeformed, so we have $\bm{U}(r, z=0)=0$.

At the boundary where the soft layer is in contact with the fluids, i.e. $z=H$,  we impose the force balance condition. In the following we introduce all the forces acting on the interface. Firstly, the capillary traction  $\bm{f}^{l}$ due to the liquid-gas surface tension is pulling the soft layer at the contact lines. To model this highly localized force, we introduce a microscopic length $\ell_m$ \citep{Hui2014,Dervaux2015,Chan2022} and a gaussian function to describe the spatial distribution of the traction. The capillary traction is given by 
\begin{eqnarray} 
\bm{f}^{l}=\gamma F(r;R,\ell_m)(-\cos\theta\hat{\bm{r}}+\sin\theta\hat{\bm{z}}),
\end{eqnarray}
where $F(r;R,\ell_m)=\exp\left[-(r-R)^2/2\ell_m^2\right]/\ell_m\sqrt{2\pi}$.
In the limit that $\ell_m\rightarrow 0$, $F(r;R,\ell_m\rightarrow 0)=\delta(r-R)$, where $\delta(r-R)$ is the Dirac delta function \cite{Karpitschka2015}. Secondly, the solid surface tension $\gamma_s$ gives a traction \citep{Snoeijer2018} 
\begin{eqnarray} 
\bm{f}^{s}=\frac{\partial \gamma_s}{\partial r}\hat{\bm{t}}+\frac{\gamma_{s}\kappa_{s}}{\vert\cos\varphi\vert}\hat{\bm{n}},
\end{eqnarray}
 where the local angle $\varphi$ of the deformed soft layer interface is defined as 
 \begin{eqnarray}\label{bcs7}
\varphi=\arctan\left(\frac{\partial u_z}{\partial r}\right),
\end{eqnarray}
with $u_z\equiv U_z(z=H)$,  
the curvature of the deformed soft layer interface $\kappa_{s}$ is given by
\begin{eqnarray}\label{bcs8}
\kappa_s=\frac{\frac{\partial^2 u_z}{\partial r^2}}{\left[1+\left(\frac{\partial u_z}{\partial r}\right)^2 \right]^{3/2}}+\frac{\frac{\partial u_z}{\partial r}}{r\left[1+\left(\frac{\partial u_z}{\partial r}\right)^2 \right]^{1/2}},
\end{eqnarray}
and $\hat{\bm{t}}\equiv \cos\varphi \hat{\bm{r}} + \sin\varphi \hat{\bm{z}}$ and $\hat{\bm{n}} \equiv -\sin\varphi \hat{\bm{r}} + \cos\varphi \hat{\bm{z}}$ are respectively  the tangential unit vector and the normal unit vector of the deformed soft layer interface. 
The factor $1/\vert\cos\varphi\vert$ is used to compensate the length change in the deformed state. Remind that our computation domain represents the undeformed state of the soft layer.
The value of solid surface tension $\gamma_s$ possesses a jump when crossing from the solid-liquid interface to the solid-gas interface. To model the jump in a continuous manner, we use the arctan function to describe the spatial change and consider that the change occurs within the microscopic length $\ell_m$, meaning that $\gamma_{s}=\gamma_{sl}+(\gamma_{sg}-\gamma_{sl})F_s(r;R,\ell_m)=\gamma_{sl}+\gamma\cos\theta_Y F_s(r;R,\ell_m)$, where $F_s(r;R,\ell_m)=\arctan\left[ (r -R )/\ell_m \right]/\pi+1/2$ .
 For soft solids, we also have to distinguish between surface energy and surface stress due to Shutterworth effect \citep{Shuttleworth1950,Andreotti2016,Masurel2019}. In this study, we neglect this effect and assume that the tension force at the solid interface is independent of the stretching of the solid. Thirdly, the Laplace pressure generated inside the droplet/bridge is pressing or pulling the solid-liquid interface.  The traction $\bm{f}^{La}$ due to the Laplace pressure is given by 
  \begin{eqnarray} 
\bm{f}^{La}=\frac{\gamma \kappa_l H_s(R- r )\left(\sin\varphi \hat{\bm{r}}-\cos\varphi\hat{\bm{z}}\right)}{\vert\cos\varphi\vert},
\end{eqnarray} \label{fLa}
where $H_s(R-r)$ is the  Heaviside step function and the curvature of the liquid-air interface is 
\begin{eqnarray}
\kappa_l=\begin{cases}
\frac{2\sin\theta}{R} & \text{for sessile droplets,}\\
-\frac{2\cos \theta}{\lambda}+\frac{1}{R} &  \text{for capillary bridges.}
\end{cases}
\end{eqnarray} \label{kappal}
Remind that for the bridge case, we assume that $R\gg \lambda$. We can see that $\kappa_l$ is positive for both hydrophilic and hydrophobic surfaces for the sessile droplet case. For the bridge case, $\kappa_l$ is negative for hydrophilic surfaces when $\cos\theta> \lambda/2R$.
 
 Fourthly,  the elastic traction due to deformation is given by 
\begin{eqnarray} 
\bm{f}^{el}=\sigma_{rz}\hat{\bm{r}}-\sigma_{zz}\hat{\bm{z}}.
\end{eqnarray}
Note that linear elasticity is used in our model, the expression of $\bm{f}^{el}$ is written with respect to the undeformed state. 

 Balancing all the tractions we have the following boundary condition at $z=H$,
\begin{eqnarray}\label{fbal}
\bm{f}^{l}+\bm{f}^{s}+\bm{f}^{La}+\bm{f}^{el}=0.
\end{eqnarray} 
To determine the contact angle $\theta$, we impose the condition that surface tensions balance each other at the contact line, which means $\bm{f}^{l}$ and $\bm{f}^{s}$ dominate over $\bm{f}^{La}$ and $\bm{f}^{el}$. Denoting the angles of the solid-liquid and solid-gas interfaces respectively as $\phi_{sl}$ and $\phi_{sg}$, see Fig. \ref{fig1}, balancing the surface tensions gives 
\begin{eqnarray}\label{stb1}
-\gamma\cos\theta-\gamma_{sl}\cos\phi_{sl}+\gamma_{sg}\cos\phi_{sg}=0
\end{eqnarray}
and
\begin{eqnarray}\label{stb2}
\gamma\sin\theta-\gamma_{sl}\sin\phi_{sl}-\gamma_{sg}\sin\phi_{sg}=0.
\end{eqnarray}
The values of $\phi_{sl}$ and $\phi_{sg}$ are determined from the profiles of the deformed solid interface. Numerically to find the solution of the contact angle $\theta$, we start with a trial value $\theta=\theta_Y$ and compute the profile of deformed solid interface to obtain $\phi_{sl}$ and $\phi_{sg}$ for the trial value of $\theta$.  We iterate the trial value of $\theta$ until the conditions (\ref{stb1}) and (\ref{stb2}) are fulfilled.

In this study, we consider that the soft material is incompressible, which means $\nu=0.5$. The incompressibility condition implies
\begin{eqnarray} \label{contin}
\nabla\cdot \bm{U}=0.
\end{eqnarray}

 Next, we non-dimensionalize the variables as the following. We introduce a length scale $l$ to rescale all the lengths and displacements. We take $l=R$ for the sessile droplet case, and $l=d$ for the capillary bridge case. The stresses are rescaled by $E$. The dimensionless lengths, displacements and stresses are: $\tilde{r}=r/l$ , $\tilde{z}=z/l$, $\tilde{H}=H/l$, $\tilde{U}_r=U_{r}/l$ , $\tilde{U}_z=U_{z}/l$,  $\tilde{p}=3p/E$ and $\tilde{\boldsymbol{\sigma}}=\boldsymbol{\sigma}/E$.  See Appendix \ref{appB} for the dimensionless governing equations and boundary conditions. Moreover, to reduce the number of control parameters, we consider only   situations where $\gamma_{sl}/\gamma=1$ and we take $\ell_m/l=10^{-7}$. We end up with the following dimensionless control parameters: 
\begin{eqnarray}
 \theta_Y,  \ \  \tilde{H}=\frac{H}{l},  \ \ S=\frac{\gamma}{El},   
\end{eqnarray}
and
\[\tilde{R}=\frac{R}{l} \  \   \  \  \text{(for capillary bridges only).}
\]
The dimensionless governing equations (\ref{ges1}), (\ref{ges2}), and (\ref{dicontin}) are solved together with the dimensionless boundary conditions  (\ref{bcs1})-(\ref{bcs6}) using the finite element method for which the details are given in Appendix \ref{appC}.

\section{Results}

\subsection{Validation: sessile droplets with $\theta_Y=90^{\circ}$ on a thick soft layer} 
\begin{figure}
\begin{center}
\includegraphics[width=0.5\textwidth]{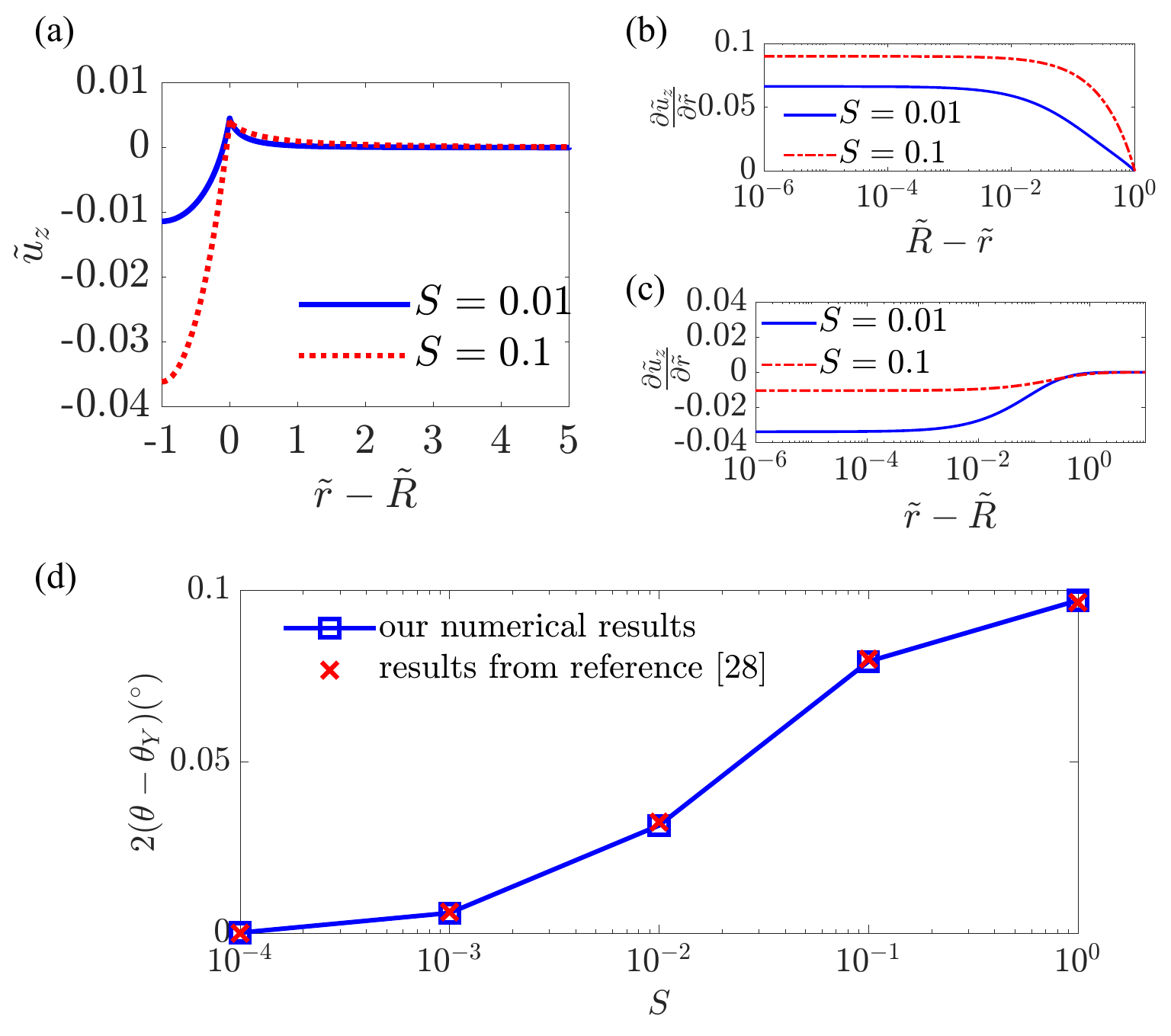}
\caption{The sessile droplet case with $\theta_Y=90^{\circ}$ and $\tilde{H}=10$. (a) The rescaled displacement $\tilde{u}_z$ as a function of $\tilde{r}-\tilde{R}$. (b) and (c): The slope $\frac{\partial\tilde{u}_z}{\partial\tilde{r}}$ as a function of $\tilde{R}-\tilde{r}$ in (b) for the solid-liquid interface  and $\tilde{r}-\tilde{R}$ in (c) for the solid-gas interface. (d) The change of contact angle $2(\theta-\theta_Y)$ as a function of the softness parameter $S$. Our numerical results is compared with  results in the reference \cite{Lubbers2014}. }\label{fig2}
\end{center}
\end{figure}

We validate our approach by comparing the contact angle change with the results in the reference \cite{Lubbers2014} where the authors study a sessile droplet with $\theta_Y=90^{\circ}$ on a semi-infinite thick soft substrate.
We compute the soft layer deformation for a sessile droplet  on a thick soft layer of $\tilde{H}=10$ and $\theta_Y=90^{\circ}$. Note that Young's angle $90^{\circ}$ implies $\gamma_{sl}=\gamma_{sg}$. Fig. \ref{fig2} (a) depicts the rescaled displacement $\tilde{u}_z\equiv \tilde{U}_z(\tilde{z}=\tilde{H})$ as a function of $\tilde{r}-\tilde{R}$ for two different values of softness parameter, i.e. $S=0.01$ and $0.1$. We can see the solid interface deforms upward to form a wetting ridge around the contact line position, i.e. $\tilde{r}-\tilde{R}=0$,   due to the pulling capillary force. Notably, although $\gamma_{sl}=\gamma_{sg}$, the interface deforms differently on the solid-liquid side and the solid-gas side. The solid-liquid interface forms a pronounced dimple resulted from the pressing by the Laplace pressure. The dimple is deeper for the softer layer of $S=0.1$. Next, to measure the angles of the interfaces at the contact line, we compute the slope of the solid interface which is shown in Fig. \ref{fig2} (b) and (c) as a function of $\tilde{R}-\tilde{r}$ (or $\tilde{r}-\tilde{R}$) in log-scale respectively for the solid-liquid side and the solid-gas side. We see when decreasing $\vert \tilde{R}-\tilde{r} \vert$, the slope reaches a plateau which indicates the range of interface where surface tensions dominate over elastic stresses \cite{Chan2022}. The angles $\phi_{sl}$ and $\phi_{sg}$ are determined from the slope at the plateau. Fig. \ref{fig2} (d) shows the derivations of $\theta$ from the Young's angle $\theta_Y$ as a function of the softness parameter $S$.  The results obtained by our method show perfect agreement with those in  the reference \cite{Lubbers2014}.

\subsection{Deformation of the soft layer for the sessile droplet case}  \label{sess}

\begin{figure}
\begin{center}
\includegraphics[width=0.5\textwidth]{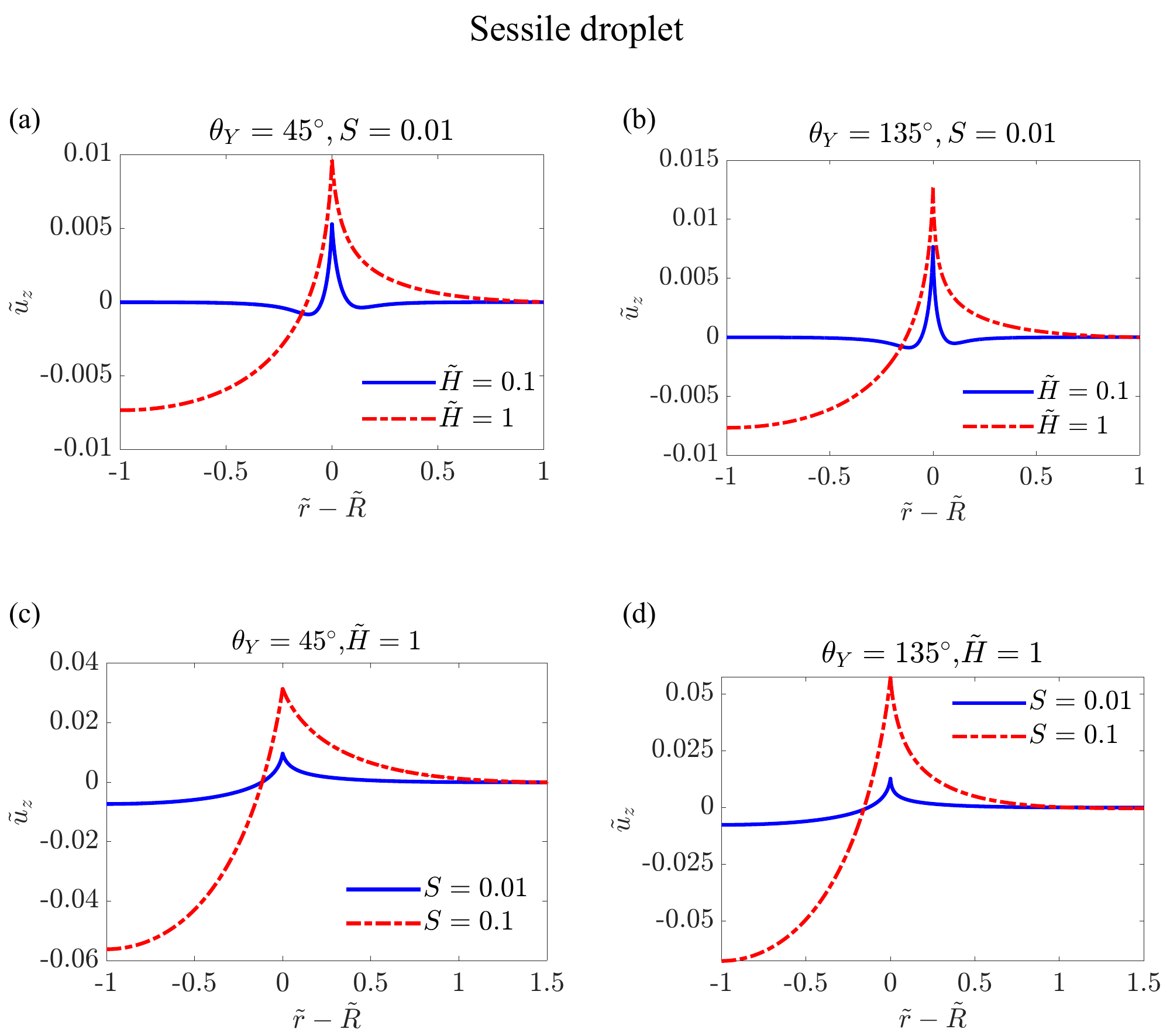}
\caption{ The rescaled displacement    $\tilde{u}_z$ as a function of $\tilde{r}-\tilde{R}$ for the sessile droplet case. In (a)  $\theta_Y=45^{\circ}$ and $S=0.01$, with different rescaled thickness of the soft layer, i.e. $\tilde{H}=0.1, 1$; (b)  $\theta_Y=135^{\circ}$ and $S=0.01$ with different rescaled thickness of the soft layer, i.e., $\tilde{H}=0.1, 1$; (c)  $\theta_Y=45^{\circ}$ and  $\tilde{H}=1$ with different softness parameter of the layer, i.e. $S=0.01, 0.1$ and (d) $\theta_Y=135^{\circ}$ and $\tilde{H}=1$ with different softness parameter of the layer, i.e. $S=0.01, 0.1$} \label{fig3}
\end{center}
\end{figure}

We examine the sessile droplet case for a hydrophilic surface of $\theta_Y=45^{\circ}$ and a hydrophobic surface of $\theta_Y=135^{\circ}$.
Figure \ref{fig3} shows the rescaled displacement  $\tilde{u}_z$ as a function of $\tilde{r}-\tilde{R} $. In Figure  \ref{fig3} (a) and (b), we fix the value of $\tilde{S}=0.01$ and compare the profiles for  $\tilde{H}=0.1$ and $\tilde{H}=1$. In \ref{fig3} (c) and (d), we compare the profiles for  $\tilde{S}=0.1$ and $\tilde{S}=1$ with a fixed value of $\tilde{H}=1$. For cases of $\tilde{H}=0.1$, which means the soft layer thickness is smaller than the droplet contact radius, we find that  a small dimple is formed on both the solid-liquid and solid-gas sides at  $\tilde{r}-\tilde{R}\approx \tilde{H}$. However, for  $\tilde{H}=1$, a dimple extends across the entire region of $0<\tilde{r}<\tilde{R}$. The asymmetry in deformation between the solid-liquid and the solid-gas sides becomes more pronounced with larger values of  $\tilde{H}$ or $\tilde{S}$. Notably, the deformation characteristics are similar for $\theta_Y=45^{\circ}$ and $\theta_Y=135^{\circ}$.

\subsection{Deformation of the soft layer for the capillary bridge case} \label{capi}

\begin{figure}
\begin{center}
\includegraphics[width=0.5\textwidth]{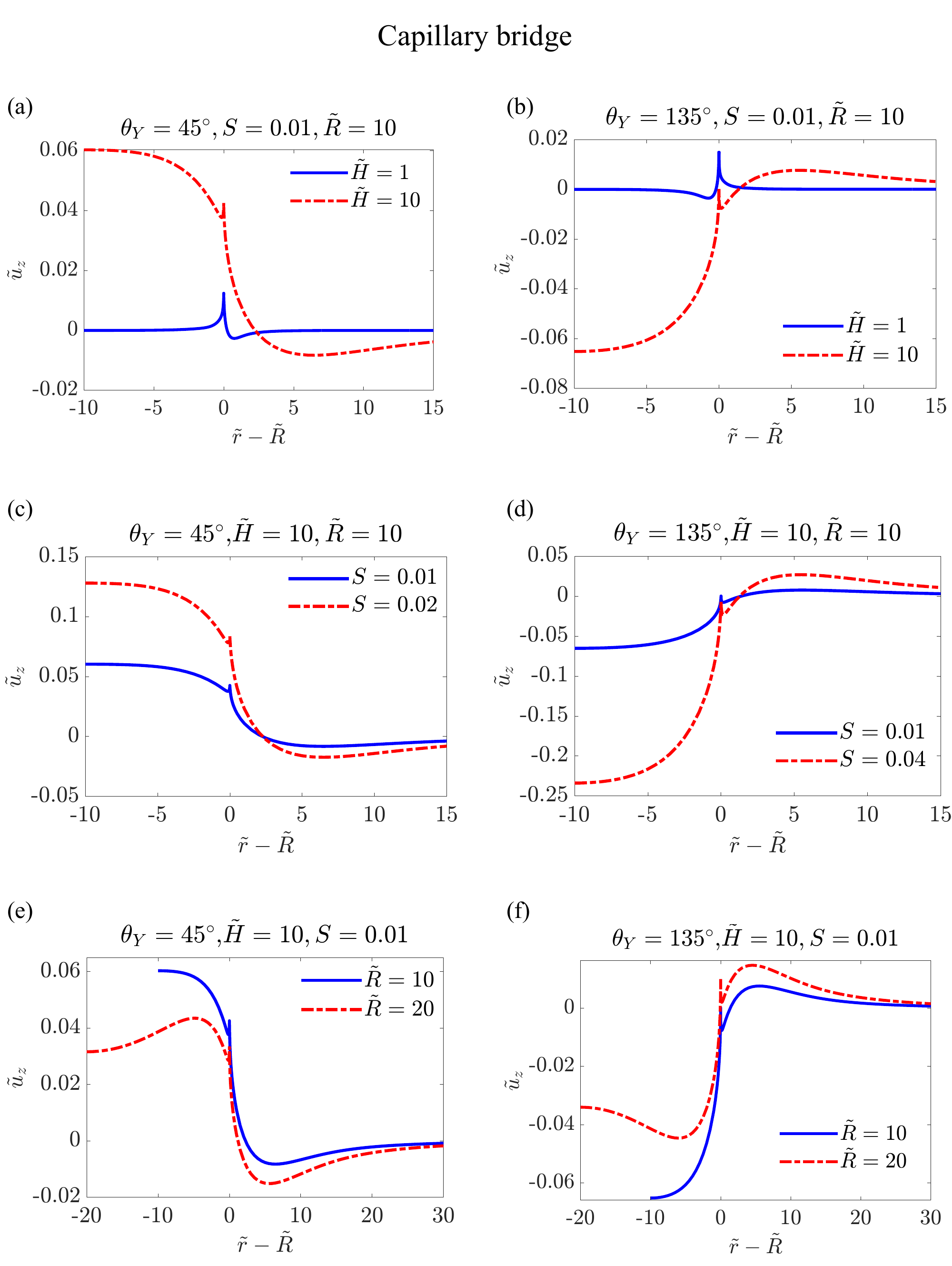}
\caption{The rescaled displacement $\tilde{u}_z$ as a function of $\tilde{r}-\tilde{R}$ for the capillary bridge case. In (a) $\theta_Y=45^{\circ}$, $S=0.01$ and $\tilde{R}=10$ with different rescaled thickness of the substrate i.e., $\tilde{H}=1, 10$; (b) $\theta_Y=135^{\circ}$, $S=0.01$ and $\tilde{R}=10$, with different rescaled thickness of the substrate i.e., $\tilde{H}=1, 10$; (c) $\theta_Y=45^{\circ}$, $\tilde{H}=1$ and $\tilde{R}=10$, with different softness parameter of the substrate i.e., $S=0.01, 0.02$ ; (d) $\theta_Y=135^{\circ}$, $\tilde{H}=1$, $\tilde{R}=10$, with different softness parameter of the substrate i.e., $S=0.01, 0.04$; (e)  $\theta_Y=45^{\circ}$, $\tilde{H}=1$ , $S=0.01$, with different rescaled radius of the capillary bridge i.e., $\tilde{R}=10, 20$ ; (d)  $\theta_Y=135^{\circ}$, $\tilde{H}=1$, $S=0.01$, with different rescaled radius of the capillary bridge i.e., $\tilde{R}=10, 20$.} \label{fig4}
\end{center}
\end{figure}

 Similarly, we examine the capillary bridge case for   $\theta_Y=45^{\circ}$ and  $\theta_Y=135^{\circ}$. Remind that all the dimensionless lengths here are rescaled by the gap separation of the two soft layers, rather than by the droplet contact radius.
 In Fig. \ref{fig4}, we plot the rescaled displacement  $\tilde{u}_z$ as a function of $\tilde{r}-\tilde{R}$.  We see that a  wetting ridge with a sharp tip forms at the contact line region, contrasting with results from  previous studies of soft wetting by capillary bridges \cite{Wexler2014,Li2014}.  Additionally, as shown in Fig. \ref{fig4}(a) and (b), increasing $\tilde{H}$ shifts the maximum value of  $\tilde{u}_z$ (denoted as $\tilde{u}_{zm}$) from the  contact line position to the solid-liquid side for $\theta_Y=45^{\circ}$  and to the solid-gas side for $\theta_Y=135^{\circ}$. This shift suggests that when the top and bottom soft layers make contact, i.e. $\tilde{u}_{zm}=0.5$, the contact point is not at the contact line position \cite{Li2014}.
 
 Compared with the sessile droplet case, we observe several key differences.  For $\theta_Y=45^{\circ}$, shown in Fig.\ref{fig4} (a), (c) and (e), the Laplace pressure pulls the solid-liquid surface, resulting in  a positive displacement for $\tilde{r}-\tilde{R}<0$. For $\theta_Y=135^{\circ}$ shown in Fig. \ref{fig4} (b), (d) and (f), the Laplace pressure  presses  the solid-liquid interface to form  a pronounced dimple, which is similar to that seen in the sessile droplet case. However, a  crest (maximum $\tilde{u}_z$) can also appear on the solid-gas side, which is not observed for the  sessile droplet case. We hypothesize that this results from a stronger Laplace pressure effect in the capillary bridge compared to the sessile droplet, relative to the pulling capillary force at the contact line. As with the sessile droplet case, the asymmetry in deformation between the solid-liquid and solid-gas sides is more pronounced with larger $\tilde{H}$ or $\tilde{S}$. In Fig. \ref{fig4}(e) and (f), we also observe that varying $\tilde{R}$ alters the deformation features, specifically shifting the positions of the maximum and minimum values of $\tilde{u}_z$.
  
\subsection{Wetting ridge rotation and the change of contact angle} \label{ridgeRot}

\begin{figure}
\begin{center}
\includegraphics[width=0.5\textwidth]{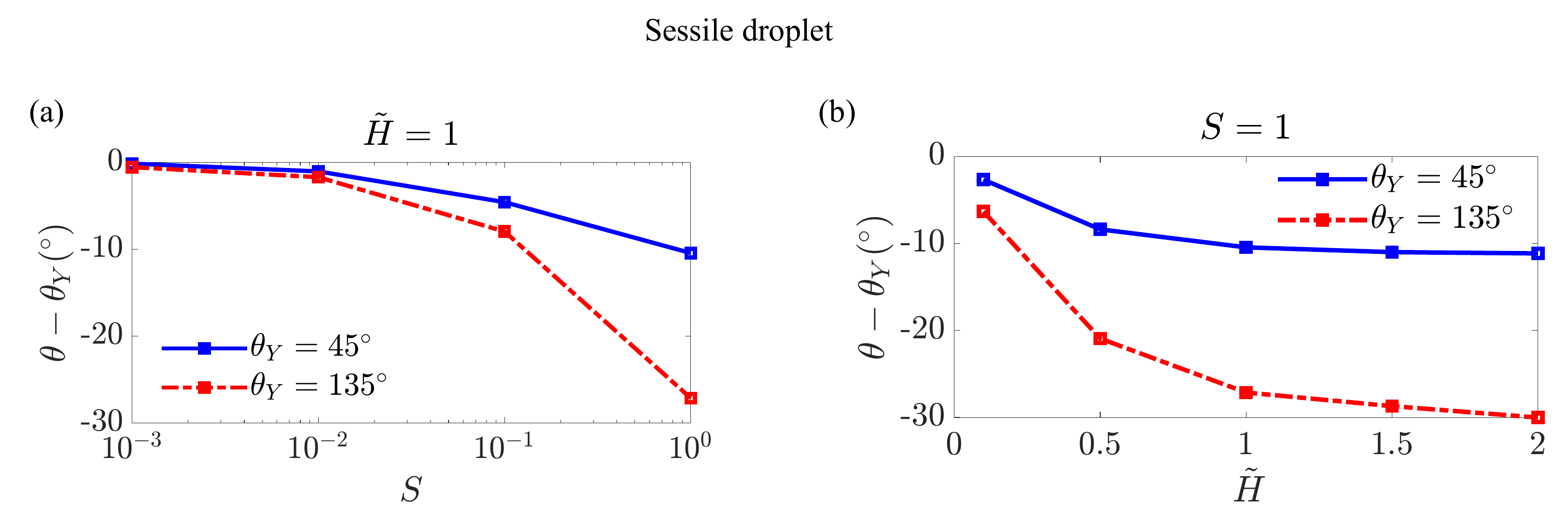}
\caption{The contact angle variation $\theta-\theta_Y$ as a function of the softness parameter $\tilde{S}$ in (a) and the rescaled soft layer thickness $\tilde{H}$ in (b) for the sessile droplet case. The blue solid line and the red dashed-dot line represent   $\theta_Y=45^{\circ}$ and  $\theta_Y=135^{\circ}$ cases, respectively. } \label{fig5}
\end{center}
\end{figure}

\begin{figure}
\begin{center}
\includegraphics[width=0.5\textwidth]{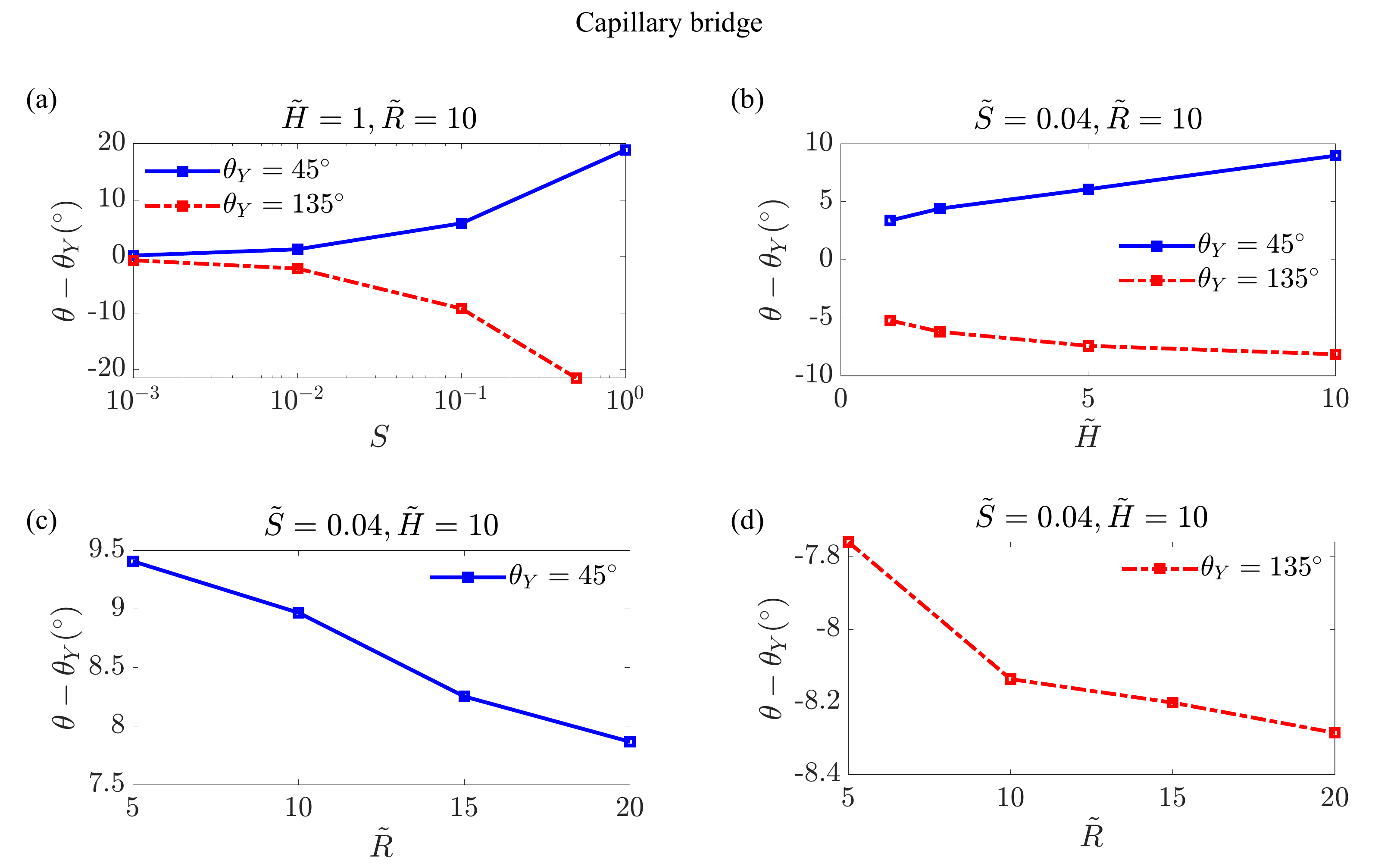}
\caption{The contact angle variation $\theta-\theta_Y$ as a function of the softness parameter $\tilde{S}$ in (a), the rescaled soft layer thickness $\tilde{H}$ in (b) and the rescaled contact radius $\tilde{R}$ in (c) and (d) for the capillary bridge case. The blue solid line and the red dashed-dot line represent   $\theta_Y=45^{\circ}$ and  $\theta_Y=135^{\circ}$ cases, respectively.} \label{fig6}
\end{center}
\end{figure}

We present how the contact angle variation, characterized  by $\theta-\theta_Y$,  changes with the control parameters for the sessile droplet case in Fig. \ref{fig5} and the capillary bridge case in Fig. \ref{fig6}. When the softness parameter $\tilde{S}\ll 1$, the contact angle remains equal to the Young's angle as shown in Fig. \ref{fig5}(a) and Fig. \ref{fig6}(a). In the sessile droplet case (Fig. \ref{fig5}), for both $\theta_Y=45^{\circ}$ and $\theta_Y=135^{\circ}$, the contact angle $\theta$ is smaller than the Young's angle, indicating a counterclockwise rotation of  the wetting ridge. The magnitude of this  rotation increases with both $S$ and $\tilde{H}$.  In contrast, for the capillary bridge case (Fig. \ref{fig6}), the contact angle $\theta$ is smaller than the Young's angle for $\theta_Y=135^{\circ}$  but larger for $\theta_Y=45^{\circ}$ . This means the wetting ridge rotates clockwise for $\theta_Y=45^{\circ}$. The increase in $\theta$ from the Young's angle  at $\theta_Y=45^{\circ}$ becomes more pronounced when enhancing $S$ or $\tilde{H}$, or with a decrease in $\tilde{R}$.

\begin{figure}
\begin{center}
\includegraphics[width=0.38\textwidth]{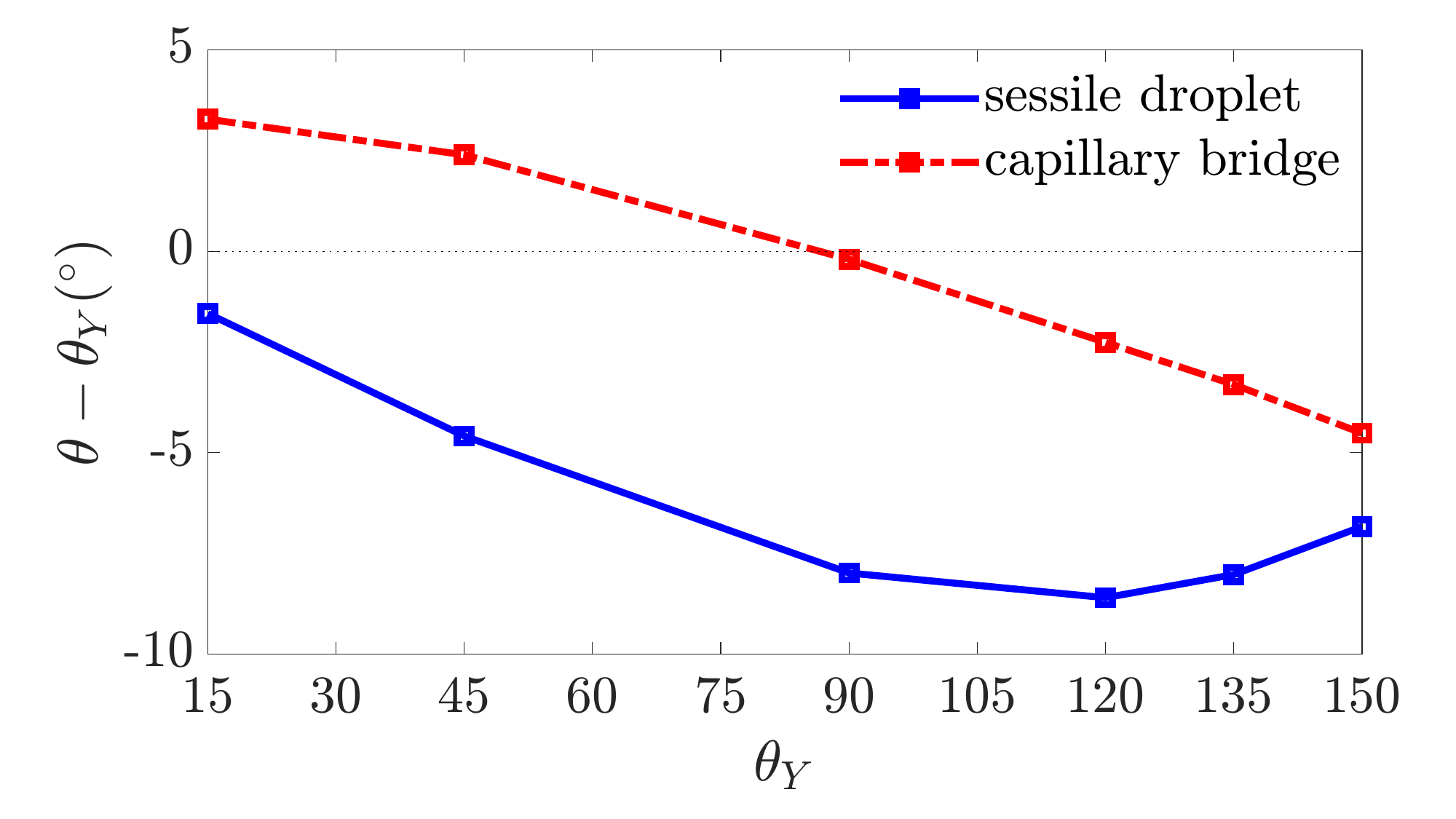}
\caption{ The contact angle variation $\theta-\theta_Y$ as a function of Young's angle $\theta_Y$. The blue line is for the sessile droplet case. Parameters:  $S=0.1$ and $\tilde{H}=1$. The red line is for the capillary bridge case. Parameters: $S=0.01$, $\tilde{H}=10$ and $\tilde{R}=10$.} \label{fig7}
\end{center}
\end{figure}

The direction of the wetting ridge rotation aligns with the sign of the Laplace pressure, $\gamma \kappa_l$, with $\kappa_l$ given by eq. (\ref{kappal}). For the sessile droplet case and the capillary bridge with $\theta_Y=135^{\circ}$, a positive Laplace pressure presses the solid-liquid interface, causing the ridge to rotate counterclockwise and reducing $\theta$. Conversely, in the capillary bridge with $\theta_Y=45^{\circ}$, a negative Laplace pressure pulls the solid-liquid interface, leading to an increase in $\theta$. To compare the contact angle change relative to the Young's angle between the sessile droplet and capillary bridge cases,  we plot $\theta- \theta_Y$ for various values of $\theta_Y$ in Fig. \ref{fig7}. Our results show that $\theta- \theta_Y$ is always negative in the sessile droplet case. The magnitude of rotation is decreasing when $\theta_Y$ approaches $0^{\circ}$ or $180^{\circ}$, where the Laplace pressure becomes negligible. Consequently, a maximum rotation occurs at an intermediate Young's angle. In the capillary bridge case, the wetting ridge rotates counterclockwise  for hydrophilic surfaces with $\theta_Y\lesssim85^{\circ}$ and clockwise  for surfaces with $\theta_Y\gtrsim 85^{\circ}$. Note that Laplace pressure vanishes when $\theta$ is slightly smaller than  $90^{\circ}$, see eq. (\ref{kappal}).

\begin{figure}
\begin{center}
\includegraphics[width=0.5\textwidth]{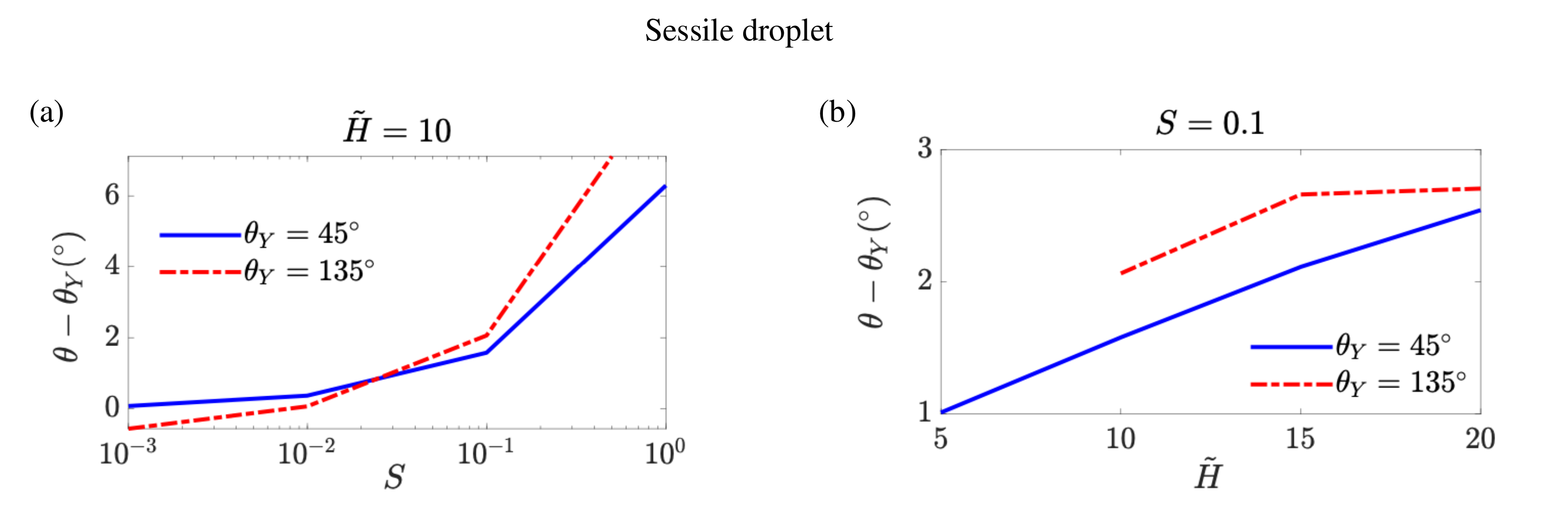}
\caption{The contact angle variation $\theta-\theta_Y$ as a function of the softness parameter $\tilde{S}$ in (a) and the rescaled thickness of the soft layer $\tilde{H}$ in (b) for the sessile droplet case when the Laplace pressure term $\bm{f}^{La}$ in Eq. \ref{fbal} is removed from our computation.  The blue solid line and the red dashed-dot line respectively represent the results for the hydrophilic surface of $\theta_Y=45^{\circ}$ and hydrophobic surface of $\theta_Y=135^{\circ}$ .} \label{fig8}
\end{center}
\end{figure}

\section{Discussions and Conclusion}
Contrary to  models that assume Young's relation for partially wetting scenarios \cite{Bostwick2014} or limit the analysis to $\theta_Y = 90^{\circ}$ \cite{Lubbers2014,Li2014,Dervaux2015}, our approach incorporates a comprehensive surface tension balance condition at the contact line, enabling it to account for a broader range of contact angle and softness of the layer. We unravel the morphology of the wetting ridge for both sessile droplets and capillary bridges. The direction of the wetting ridge rotation critically depends on the sign of the Laplace pressure. For example for the hydrophilic capillary bridge case, a negative Laplace pressure pulls the solid-liquid interface and the contact angle is enhanced from the Young's angle.
Assuming that a droplet migrates from regions of higher to lower  surface energy, a hydrophilic capillary bridge  is expected to move from thicker regions of a soft layer toward thinner areas, in contrast to the behavior observed for a sessile droplet. 

In a numerical study by  Bueno et al.  \cite{Bueno2018} employing a nonlinear elasticity model with a diffuse interface, it is shown that sessile droplets on hydrophilic surfaces migrate toward thicker regions of a soft layer, consistent with experimental observations by Style et al.  \cite{Style2013pnas}. Remarkably, on hydrophobic surfaces, the droplets  move in an opposite direction, namely from   thicker regions  to  thinner regions. The study further reveals that on such surfaces, the contact angle becomes larger when the layer is softer or thicker. The authors point out that although Laplace pressure rotates the ridge toward the droplet side, resulting in a reduced  contact angle, the capillary force at the contact line tends to rotate the ridge in the opposite direction when the surface is hydrophobic. This effect dominates when  $\theta_Y$ is above a critical value. In contrast, our results show that the wetting ridge rotates in the same direction for all values of $\theta_Y$ as illustrated in Fig. \ref{fig7} for the sessile droplet case. Interestingly, we observe that the maximum magnitude of the contact angle variation occurs when the surface is hydrophobic.  Given the differences in model assumptions, such as nonlinearity and boundary conditions at the contact line, further investigation is needed to reconcile these discrepancies. 

To further explore this, we examine a specific situation where the Laplace pressure term $\bm{f}^{La}$ in Eq. \ref{fbal} is removed from our computation, leaving only the capillary pulling force at the contact line acting on the soft layer. In Fig. \ref{fig8}, we show the resulting contact angle variation for the sessile droplet case. We find that the capillary pulling force alone rotates the ridge such that the contact angle is enhanced as  $\tilde{S}$ or $\tilde{H}$ increases for both the hydrophilic surface of $\theta_Y=45^{\circ}$ and hydrophobic surface of $\theta_Y=135^{\circ}$. This rotation direction is opposite to that induced by the Laplace pressure.  Nevertheless, as our complete analysis indicates, the direction of ridge rotation is still governed by the sign of the Laplace pressure. 
Future experimental studies will be crucial for clarifying the physical mechanisms underlying these behaviors.

The complex interplay between soft layer deformation and droplet contact angle remains far from fully understood, particularly in confined geometries beyond the classic sessile droplet case. Extensive theoretical and experimental investigations are needed to deepen our understanding in this area, providing valuable insights for controlling droplet motion through elastocapillarity.

\section{Appendix:}

\subsection{Appendix A: Relations between the components of the stress tensor and the displacements in cylindrical coordinates} \label{appA}
 \begin{eqnarray} \label{elst1}
\sigma_{rr} = -p+\frac{E}{(1+\nu)}  \left(\frac{{\partial}U_{r}}{{\partial}r} -\frac{1}{3}\nabla\cdot\bm{U}\right),
\end{eqnarray}

\begin{eqnarray} \label{elst2}
\sigma_{zz} = -p+\frac{E}{(1+\nu)}  \left(\frac{{\partial}U_{z}}{{\partial}z} -\frac{1}{3}\nabla\cdot\bm{U}\right),
\end{eqnarray}

\begin{eqnarray} \label{elst3}
\sigma_{\phi\phi} = -p+\frac{E}{(1+\nu)}  \left(\frac{U_{r}}{r} -\frac{1}{3}\nabla\cdot\bm{U}\right),
\end{eqnarray}

\begin{eqnarray} \label{elst4}\sigma_{rz} = \frac{E}{(1+\nu)}  \left(\frac{{\partial}U_{r}}{{\partial}z} +\frac{{\partial}U_{z}}{{\partial}r}\right),
\end{eqnarray} 
where the isotropic part of the stress tensor (or the pressure)
\begin{eqnarray} \label{elst5}
p= -\frac{E}{3(1-2\nu)} \nabla\cdot\bm{U}.
\end{eqnarray}

\subsection{Appendix B: The dimensionless governing equations and boundary conditions}\label{appB}

For the elastic deformation,  the dimensionless form of $\bigtriangledown\cdot \bm{\sigma}=0$ is
\begin{eqnarray} \label{ges1}
\tilde{\nabla}^{2}\tilde{U}_r - \frac{\tilde{U}_r}{\tilde{r}^{2}} -\frac{\partial \tilde{p}}{{\partial}\tilde{r}}=0
\end{eqnarray} 
in $r$-direction and
\begin{eqnarray} \label{ges2}
\tilde{\nabla}^{2}\tilde{U}_z - \frac{\partial \tilde{p}}{{\partial}\tilde{z}}=0
\end{eqnarray} 
in $z$-direction.
The dimensionless form of the incompressibility condition (eq. \ref{contin}) is 
\begin{eqnarray} \label{dicontin}
\tilde{\nabla}\cdot \tilde{\bm{U}}=0.
\end{eqnarray}

The boundary conditions far away from the droplet at $\tilde{r}= \tilde{L}$ and at the soft/rigid solid interface respectively are
\begin{eqnarray}\label{bcs1}
\tilde{\bm{U}}(\tilde{r}=\tilde{L}, \tilde{z})=0
\end{eqnarray} 
and
\begin{eqnarray}\label{bcs2}
\tilde{\bm{U}}(\tilde{r}, \tilde{z}=0)=0.
\end{eqnarray} 

At $\tilde{r}=0$, the symmetry property gives 
\begin{eqnarray}\label{bcs3}
\tilde{U_r}(\tilde{r}=0, \tilde{z})=0,
\end{eqnarray} 
and
\begin{eqnarray}\label{bcs4}
\frac{\partial \tilde{U_z}}{\partial \tilde{r}}(\tilde{r}=0, \tilde{z})=0.
\end{eqnarray}

At the soft solid/fluid interface $\tilde{z}=\tilde{H}$,  the force balance condition (eq. \ref{fbal}) for the $r$-components and $z$-components respectively gives 
\begin{eqnarray}\label{bcs5}
\frac{\tilde{\sigma}_{rz}}{S} +\tilde{\kappa}_lH_s(\tilde{R}-\tilde{r}) \frac{\sin\varphi}{\vert\cos\varphi\vert}-\cos\theta \tilde{F} \nonumber\\
-\tilde{\gamma}_{s}\tilde{\kappa}_s \frac{\sin\varphi}{\vert\cos\varphi\vert}\nonumber\\
+ \frac{\partial \tilde{\gamma}_s}{\partial \tilde{r}} \cos\varphi
 =0.
\end{eqnarray}
 and
\begin{eqnarray}\label{bcs6}
- \frac{\tilde{\sigma}_{zz}}{S}
-\tilde{\kappa}_lH_s(\tilde{R}-\tilde{r}) \sign(\cos\varphi)+\sin\theta \tilde{F}\nonumber\\
+\tilde{\gamma}_{s}\tilde{\kappa}_s \sign(\cos\varphi)\nonumber\\
+ \frac{\partial \tilde{\gamma}_s}{\partial \tilde{r}} \sin\varphi
=0
\end{eqnarray}
where  $\tilde{\kappa}_l={\kappa}_ll$, $\tilde{\kappa}_s={\kappa}_sl$, $\tilde{\gamma}_{s}=\frac{\gamma_{s}}{\gamma}$, and
$\tilde{F}=\exp\left[-(\tilde{r}-\tilde{R})^2/2\tilde{\ell}_m^2\right]/\tilde{\ell}_m\sqrt{2\pi}$.

%

\subsection{Appendix C: Finite element method}\label{appC}

The rescaled displacement $\tilde{\bm{U}}$ is computed by solving the governing equations (\ref{ges1})-(\ref{dicontin}) together with the boundary conditions  (\ref{bcs1})- (\ref{bcs6}) by using a finite element method (FEM) with a Newton solver from the FEM library FEniCS\cite{logg2012automated}. We have used the adaptive mesh sizing such that the mesh size far away from the contact line is chosen to ensure the change of $|\theta-\theta_Y|$ is less than 2$\%$. The smallest mesh size in the contact line region is 1$\%$ of  $l_m$.

\begin{figure}
\begin{center}
\includegraphics[width=0.5\textwidth]{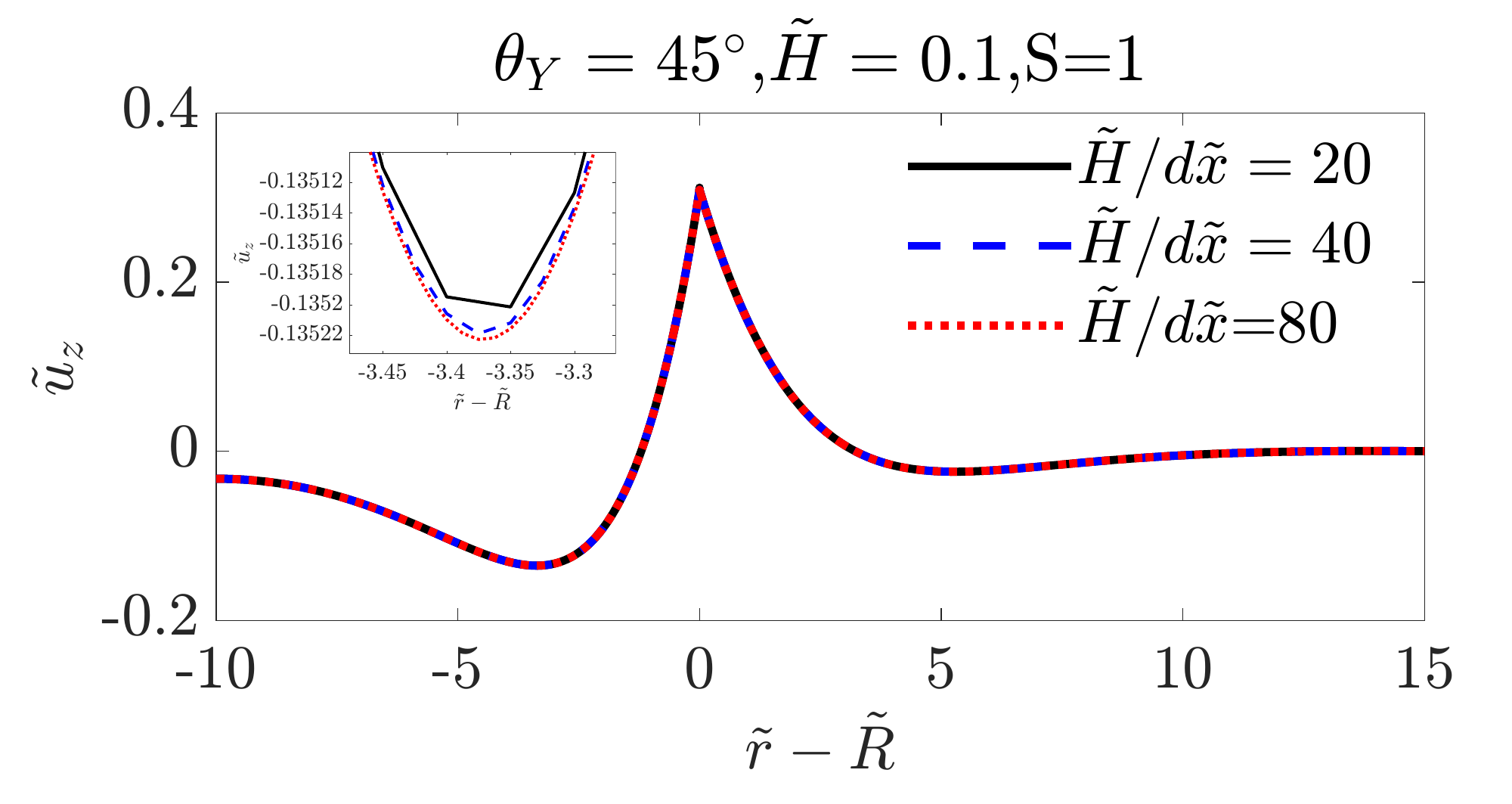}
\caption{ The rescaled displacement $\tilde{u}_z$ as a function of $\tilde{r}-\tilde{R}$ for 3 different mesh size $\tilde{H}/d\tilde{x}_0$ is the mesh size far away from the contact line.We show the numerical results with 3 different mesh sizes.The inset: the zoom into the dimple position.}  \label{fig9}
\end{center}
\end{figure}

  Mesh convergence of the numerical solver is demonstrated for the sessile droplet case shown in Fig \ref{fig9} in which  $\tilde{u}_z$ is plotted as a function of $\tilde{r}-\tilde{R}$ with three mesh resolutions: $\tilde{H}/d\tilde{x}=20$, $40$ and $80$, where $d\tilde{x}$ is the mesh size far away from the contact line. The inset shows the zoom into the dimple position. The plot demonstrates good mesh convergence.

\section*{Conflicts of interest}
There are no conflicts to declare.

\section*{Acknowledgements}
BXZ and TSC gratefully acknowledge financial support from the Research Council of Norway (Project No. 315110). 



\balance


\bibliographystyle{rsc} 

\bibliography{Ridge_form}

\providecommand*{\mcitethebibliography}{\thebibliography}
\csname @ifundefined\endcsname{endmcitethebibliography}
{\let\endmcitethebibliography\endthebibliography}{}
\begin{mcitethebibliography}{45}
\providecommand*{\natexlab}[1]{#1}
\providecommand*{\mciteSetBstSublistMode}[1]{}
\providecommand*{\mciteSetBstMaxWidthForm}[2]{}
\providecommand*{\mciteBstWouldAddEndPuncttrue}
  {\def\EndOfBibitem{\unskip.}}
\providecommand*{\mciteBstWouldAddEndPunctfalse}
  {\let\EndOfBibitem\relax}
\providecommand*{\mciteSetBstMidEndSepPunct}[3]{}
\providecommand*{\mciteSetBstSublistLabelBeginEnd}[3]{}
\providecommand*{\EndOfBibitem}{}
\mciteSetBstSublistMode{f}
\mciteSetBstMaxWidthForm{subitem}
{(\emph{\alph{mcitesubitemcount}})}
\mciteSetBstSublistLabelBeginEnd{\mcitemaxwidthsubitemform\space}
{\relax}{\relax}

\bibitem[Tenjimbayashi and Manabe(2022)]{Tenjimbayashi2022}
M.~Tenjimbayashi and K.~Manabe, \emph{A review on control of droplet motion based on wettability modulation: principles, design strategies, recent progress, and applications}, 2022\relax
\mciteBstWouldAddEndPuncttrue
\mciteSetBstMidEndSepPunct{\mcitedefaultmidpunct}
{\mcitedefaultendpunct}{\mcitedefaultseppunct}\relax
\EndOfBibitem
\bibitem[Pericet-C{\'{a}}mara \emph{et~al.}(2008)Pericet-C{\'{a}}mara, Best, Butt, and Bonaccurso]{Pericet-Camara2008}
R.~Pericet-C{\'{a}}mara, A.~Best, H.~J. Butt and E.~Bonaccurso, \emph{Langmuir}, 2008, \textbf{24}, 10565--10568\relax
\mciteBstWouldAddEndPuncttrue
\mciteSetBstMidEndSepPunct{\mcitedefaultmidpunct}
{\mcitedefaultendpunct}{\mcitedefaultseppunct}\relax
\EndOfBibitem
\bibitem[Jerison \emph{et~al.}(2011)Jerison, Xu, Wilen, and Dufresne]{Jerison2011}
E.~R. Jerison, Y.~Xu, L.~A. Wilen and E.~R. Dufresne, \emph{Physical Review Letters}, 2011, \textbf{106}, 1--4\relax
\mciteBstWouldAddEndPuncttrue
\mciteSetBstMidEndSepPunct{\mcitedefaultmidpunct}
{\mcitedefaultendpunct}{\mcitedefaultseppunct}\relax
\EndOfBibitem
\bibitem[Limat(2012)]{Limat2012}
L.~Limat, \emph{The European Physical Journal E}, 2012, \textbf{35}, 1--13\relax
\mciteBstWouldAddEndPuncttrue
\mciteSetBstMidEndSepPunct{\mcitedefaultmidpunct}
{\mcitedefaultendpunct}{\mcitedefaultseppunct}\relax
\EndOfBibitem
\bibitem[Yu(2012)]{Yu2012}
Y.~S. Yu, \emph{Applied Mathematics and Mechanics (English Edition)}, 2012, \textbf{33}, 1095--1114\relax
\mciteBstWouldAddEndPuncttrue
\mciteSetBstMidEndSepPunct{\mcitedefaultmidpunct}
{\mcitedefaultendpunct}{\mcitedefaultseppunct}\relax
\EndOfBibitem
\bibitem[Style and Dufresne(2012)]{Style2012}
R.~W. Style and E.~R. Dufresne, \emph{Soft Matter}, 2012, \textbf{8}, 7177--7184\relax
\mciteBstWouldAddEndPuncttrue
\mciteSetBstMidEndSepPunct{\mcitedefaultmidpunct}
{\mcitedefaultendpunct}{\mcitedefaultseppunct}\relax
\EndOfBibitem
\bibitem[Style \emph{et~al.}(2013)Style, Boltyanskiy, Che, Wettlaufer, Wilen, and Dufresne]{Style2013prl}
R.~W. Style, R.~Boltyanskiy, Y.~Che, J.~S. Wettlaufer, L.~A. Wilen and E.~R. Dufresne, \emph{Physical Review Letters}, 2013, \textbf{110}, 1--5\relax
\mciteBstWouldAddEndPuncttrue
\mciteSetBstMidEndSepPunct{\mcitedefaultmidpunct}
{\mcitedefaultendpunct}{\mcitedefaultseppunct}\relax
\EndOfBibitem
\bibitem[Kajiya \emph{et~al.}(2013)Kajiya, Daerr, Narita, Royon, Lequeux, and Limat]{Kajiya2013}
T.~Kajiya, A.~Daerr, T.~Narita, L.~Royon, F.~Lequeux and L.~Limat, \emph{Soft Matter}, 2013, \textbf{9}, 454--461\relax
\mciteBstWouldAddEndPuncttrue
\mciteSetBstMidEndSepPunct{\mcitedefaultmidpunct}
{\mcitedefaultendpunct}{\mcitedefaultseppunct}\relax
\EndOfBibitem
\bibitem[Park \emph{et~al.}(2014)Park, Weon, Lee, Lee, Kim, and Je]{Park2014}
S.~J. Park, B.~M. Weon, J.~S. Lee, J.~Lee, J.~Kim and J.~H. Je, \emph{Nature Communications}, 2014, \textbf{5}, 1--7\relax
\mciteBstWouldAddEndPuncttrue
\mciteSetBstMidEndSepPunct{\mcitedefaultmidpunct}
{\mcitedefaultendpunct}{\mcitedefaultseppunct}\relax
\EndOfBibitem
\bibitem[Bostwick \emph{et~al.}(2014)Bostwick, Shearer, and Daniels]{Bostwick2014}
J.~B. Bostwick, M.~Shearer and K.~E. Daniels, \emph{Soft Matter}, 2014, \textbf{10}, 7361--7369\relax
\mciteBstWouldAddEndPuncttrue
\mciteSetBstMidEndSepPunct{\mcitedefaultmidpunct}
{\mcitedefaultendpunct}{\mcitedefaultseppunct}\relax
\EndOfBibitem
\bibitem[Karpitschka \emph{et~al.}(2015)Karpitschka, Das, {Van Gorcum}, Perrin, Andreotti, and Snoeijer]{Karpitschka2015}
S.~Karpitschka, S.~Das, M.~{Van Gorcum}, H.~Perrin, B.~Andreotti and J.~H. Snoeijer, \emph{Nature Communications}, 2015, \textbf{6}, 1--7\relax
\mciteBstWouldAddEndPuncttrue
\mciteSetBstMidEndSepPunct{\mcitedefaultmidpunct}
{\mcitedefaultendpunct}{\mcitedefaultseppunct}\relax
\EndOfBibitem
\bibitem[Style \emph{et~al.}(2017)Style, Jagota, Hui, and Dufresne]{Style2017}
R.~W. Style, A.~Jagota, C.~Y. Hui and E.~R. Dufresne, \emph{Annual Review of Condensed Matter Physics}, 2017, \textbf{8}, 99--118\relax
\mciteBstWouldAddEndPuncttrue
\mciteSetBstMidEndSepPunct{\mcitedefaultmidpunct}
{\mcitedefaultendpunct}{\mcitedefaultseppunct}\relax
\EndOfBibitem
\bibitem[Fernandez-Toledano \emph{et~al.}(2017)Fernandez-Toledano, Blake, Lambert, and {De Coninck}]{Fernandez-Toledano2017}
J.~C. Fernandez-Toledano, T.~D. Blake, P.~Lambert and J.~{De Coninck}, \emph{Advances in Colloid and Interface Science}, 2017, \textbf{245}, 102--107\relax
\mciteBstWouldAddEndPuncttrue
\mciteSetBstMidEndSepPunct{\mcitedefaultmidpunct}
{\mcitedefaultendpunct}{\mcitedefaultseppunct}\relax
\EndOfBibitem
\bibitem[Andreotti and Snoeijer(2020)]{Andreotti2020}
B.~Andreotti and J.~H. Snoeijer, \emph{Annual Review of Fluid Mechanics}, 2020, \textbf{52}, 285--308\relax
\mciteBstWouldAddEndPuncttrue
\mciteSetBstMidEndSepPunct{\mcitedefaultmidpunct}
{\mcitedefaultendpunct}{\mcitedefaultseppunct}\relax
\EndOfBibitem
\bibitem[Pandey \emph{et~al.}(2020)Pandey, Andreotti, Karpitschka, {Van Zwieten}, {Van Brummelen}, and Snoeijer]{Pandey2020}
A.~Pandey, B.~Andreotti, S.~Karpitschka, G.~J. {Van Zwieten}, E.~H. {Van Brummelen} and J.~H. Snoeijer, \emph{Physical Review X}, 2020, \textbf{10}, 31067\relax
\mciteBstWouldAddEndPuncttrue
\mciteSetBstMidEndSepPunct{\mcitedefaultmidpunct}
{\mcitedefaultendpunct}{\mcitedefaultseppunct}\relax
\EndOfBibitem
\bibitem[Chan(2022)]{Chan2022}
T.~S. Chan, \emph{Soft Matter}, 2022,  7280--7290\relax
\mciteBstWouldAddEndPuncttrue
\mciteSetBstMidEndSepPunct{\mcitedefaultmidpunct}
{\mcitedefaultendpunct}{\mcitedefaultseppunct}\relax
\EndOfBibitem
\bibitem[Zheng \emph{et~al.}(2023)Zheng, Pedersen, Carlson, and Chan]{Zheng2023}
B.~X. Zheng, C.~Pedersen, A.~Carlson and T.~S. Chan, \emph{Soft Matter}, 2023, \textbf{19}, 8988--8996\relax
\mciteBstWouldAddEndPuncttrue
\mciteSetBstMidEndSepPunct{\mcitedefaultmidpunct}
{\mcitedefaultendpunct}{\mcitedefaultseppunct}\relax
\EndOfBibitem
\bibitem[Style \emph{et~al.}(2013)Style, Che, Park, Weon, Je, Hyland, German, Power, Wilen, Wettlaufer, and Dufresne]{Style2013pnas}
R.~W. Style, Y.~Che, S.~J. Park, B.~M. Weon, J.~H. Je, C.~Hyland, G.~K. German, M.~P. Power, L.~A. Wilen, J.~S. Wettlaufer and E.~R. Dufresne, \emph{Proceedings of the National Academy of Sciences of the United States of America}, 2013, \textbf{110}, 12541--12544\relax
\mciteBstWouldAddEndPuncttrue
\mciteSetBstMidEndSepPunct{\mcitedefaultmidpunct}
{\mcitedefaultendpunct}{\mcitedefaultseppunct}\relax
\EndOfBibitem
\bibitem[Liu \emph{et~al.}(2017)Liu, Nadermann, He, Strogatz, Hui, and Jagota]{Liu2017}
T.~Liu, N.~Nadermann, Z.~He, S.~H. Strogatz, C.~Y. Hui and A.~Jagota, \emph{Langmuir}, 2017, \textbf{33}, 4942--4947\relax
\mciteBstWouldAddEndPuncttrue
\mciteSetBstMidEndSepPunct{\mcitedefaultmidpunct}
{\mcitedefaultendpunct}{\mcitedefaultseppunct}\relax
\EndOfBibitem
\bibitem[Bueno \emph{et~al.}(2018)Bueno, Bazilevs, Juanes, and Gomez]{Bueno2018}
J.~Bueno, Y.~Bazilevs, R.~Juanes and H.~Gomez, \emph{Soft Matter}, 2018, \textbf{14}, 1417--1426\relax
\mciteBstWouldAddEndPuncttrue
\mciteSetBstMidEndSepPunct{\mcitedefaultmidpunct}
{\mcitedefaultendpunct}{\mcitedefaultseppunct}\relax
\EndOfBibitem
\bibitem[Bardall \emph{et~al.}(2020)Bardall, Chen, Daniels, and Shearer]{Bardall2020}
A.~Bardall, S.~Y. Chen, K.~E. Daniels and M.~Shearer, \emph{IMA Journal of Applied Mathematics (Institute of Mathematics and Its Applications)}, 2020, \textbf{85}, 495--512\relax
\mciteBstWouldAddEndPuncttrue
\mciteSetBstMidEndSepPunct{\mcitedefaultmidpunct}
{\mcitedefaultendpunct}{\mcitedefaultseppunct}\relax
\EndOfBibitem
\bibitem[Gomez and Velay-Lizancos(2020)]{Gomez2020}
H.~Gomez and M.~Velay-Lizancos, \emph{European Physical Journal: Special Topics}, 2020, \textbf{229}, 265--273\relax
\mciteBstWouldAddEndPuncttrue
\mciteSetBstMidEndSepPunct{\mcitedefaultmidpunct}
{\mcitedefaultendpunct}{\mcitedefaultseppunct}\relax
\EndOfBibitem
\bibitem[Henkel \emph{et~al.}(2021)Henkel, Snoeijer, and Thiele]{Henkel2021}
C.~Henkel, J.~H. Snoeijer and U.~Thiele, \emph{Soft Matter}, 2021, \textbf{17}, 10359--10375\relax
\mciteBstWouldAddEndPuncttrue
\mciteSetBstMidEndSepPunct{\mcitedefaultmidpunct}
{\mcitedefaultendpunct}{\mcitedefaultseppunct}\relax
\EndOfBibitem
\bibitem[Tamim and Bostwick(2021)]{Tamim2021}
S.~I. Tamim and J.~B. Bostwick, \emph{Physical Review E}, 2021, \textbf{104}, 1--10\relax
\mciteBstWouldAddEndPuncttrue
\mciteSetBstMidEndSepPunct{\mcitedefaultmidpunct}
{\mcitedefaultendpunct}{\mcitedefaultseppunct}\relax
\EndOfBibitem
\bibitem[Kajouri \emph{et~al.}(2024)Kajouri, Theodorakis, and Milchev]{Kajouri2024}
R.~Kajouri, P.~E. Theodorakis and A.~Milchev, \emph{Langmuir}, 2024, \textbf{40}, 17779--17785\relax
\mciteBstWouldAddEndPuncttrue
\mciteSetBstMidEndSepPunct{\mcitedefaultmidpunct}
{\mcitedefaultendpunct}{\mcitedefaultseppunct}\relax
\EndOfBibitem
\bibitem[Yu and Zhao(2009)]{Yu2009}
Y.~S. Yu and Y.~P. Zhao, \emph{Journal of Colloid and Interface Science}, 2009, \textbf{339}, 489--494\relax
\mciteBstWouldAddEndPuncttrue
\mciteSetBstMidEndSepPunct{\mcitedefaultmidpunct}
{\mcitedefaultendpunct}{\mcitedefaultseppunct}\relax
\EndOfBibitem
\bibitem[Snoeijer \emph{et~al.}(2018)Snoeijer, Rolley, and Andreotti]{Snoeijer2018}
J.~H. Snoeijer, E.~Rolley and B.~Andreotti, \emph{Physical Review Letters}, 2018, \textbf{121}, 68003\relax
\mciteBstWouldAddEndPuncttrue
\mciteSetBstMidEndSepPunct{\mcitedefaultmidpunct}
{\mcitedefaultendpunct}{\mcitedefaultseppunct}\relax
\EndOfBibitem
\bibitem[Lubbers \emph{et~al.}(2014)Lubbers, Weijs, Botto, Das, Andreotti, and Snoeijer]{Lubbers2014}
L.~A. Lubbers, J.~H. Weijs, L.~Botto, S.~Das, B.~Andreotti and J.~H. Snoeijer, \emph{Journal of Fluid Mechanics}, 2014, \textbf{747}, R1\relax
\mciteBstWouldAddEndPuncttrue
\mciteSetBstMidEndSepPunct{\mcitedefaultmidpunct}
{\mcitedefaultendpunct}{\mcitedefaultseppunct}\relax
\EndOfBibitem
\bibitem[Dervaux and Limat(2015)]{Dervaux2015}
J.~Dervaux and L.~Limat, \emph{Proceedings of the Royal Society A: Mathematical, Physical and Engineering Sciences}, 2015, \textbf{471}, year\relax
\mciteBstWouldAddEndPuncttrue
\mciteSetBstMidEndSepPunct{\mcitedefaultmidpunct}
{\mcitedefaultendpunct}{\mcitedefaultseppunct}\relax
\EndOfBibitem
\bibitem[Fortes(1982)]{FORTES1982}
M.~Fortes, \emph{Journal of Colloid and Interface Science}, 1982, \textbf{88}, 338--352\relax
\mciteBstWouldAddEndPuncttrue
\mciteSetBstMidEndSepPunct{\mcitedefaultmidpunct}
{\mcitedefaultendpunct}{\mcitedefaultseppunct}\relax
\EndOfBibitem
\bibitem[Carter(1988)]{CARTER1988}
W.~Carter, \emph{Acta Metallurgica}, 1988, \textbf{36}, 2283--2292\relax
\mciteBstWouldAddEndPuncttrue
\mciteSetBstMidEndSepPunct{\mcitedefaultmidpunct}
{\mcitedefaultendpunct}{\mcitedefaultseppunct}\relax
\EndOfBibitem
\bibitem[Marmur(1993)]{Marmur1993}
A.~Marmur, \emph{Langmuir}, 1993, \textbf{9}, 1922--1926\relax
\mciteBstWouldAddEndPuncttrue
\mciteSetBstMidEndSepPunct{\mcitedefaultmidpunct}
{\mcitedefaultendpunct}{\mcitedefaultseppunct}\relax
\EndOfBibitem
\bibitem[Herminghaus(2005)]{Herminghaus2005}
S.~Herminghaus, \emph{Advances in Physics}, 2005, \textbf{54}, 221--261\relax
\mciteBstWouldAddEndPuncttrue
\mciteSetBstMidEndSepPunct{\mcitedefaultmidpunct}
{\mcitedefaultendpunct}{\mcitedefaultseppunct}\relax
\EndOfBibitem
\bibitem[Rabinovich \emph{et~al.}(2005)Rabinovich, Esayanur, and Moudgil]{Rabinovich2005}
Y.~I. Rabinovich, M.~S. Esayanur and B.~M. Moudgil, \emph{Langmuir}, 2005, \textbf{21}, 10992--10997\relax
\mciteBstWouldAddEndPuncttrue
\mciteSetBstMidEndSepPunct{\mcitedefaultmidpunct}
{\mcitedefaultendpunct}{\mcitedefaultseppunct}\relax
\EndOfBibitem
\bibitem[Qian and Gao(2006)]{Qian2006}
J.~Qian and H.~Gao, \emph{Acta Biomaterialia}, 2006, \textbf{2}, 51--58\relax
\mciteBstWouldAddEndPuncttrue
\mciteSetBstMidEndSepPunct{\mcitedefaultmidpunct}
{\mcitedefaultendpunct}{\mcitedefaultseppunct}\relax
\EndOfBibitem
\bibitem[{De Souza} \emph{et~al.}(2008){De Souza}, Brinkmann, Mohrdieck, and Arzt]{DeSouza2008}
E.~J. {De Souza}, M.~Brinkmann, C.~Mohrdieck and E.~Arzt, \emph{Langmuir}, 2008, \textbf{24}, 8813--8820\relax
\mciteBstWouldAddEndPuncttrue
\mciteSetBstMidEndSepPunct{\mcitedefaultmidpunct}
{\mcitedefaultendpunct}{\mcitedefaultseppunct}\relax
\EndOfBibitem
\bibitem[Butt and Kappl(2009)]{Butt2009}
H.~J. Butt and M.~Kappl, \emph{Advances in Colloid and Interface Science}, 2009, \textbf{146}, 48--60\relax
\mciteBstWouldAddEndPuncttrue
\mciteSetBstMidEndSepPunct{\mcitedefaultmidpunct}
{\mcitedefaultendpunct}{\mcitedefaultseppunct}\relax
\EndOfBibitem
\bibitem[Chen \emph{et~al.}(2013)Chen, Amirfazli, and Tang]{Chen2013}
H.~Chen, A.~Amirfazli and T.~Tang, \emph{Langmuir}, 2013, \textbf{29}, 3310--3319\relax
\mciteBstWouldAddEndPuncttrue
\mciteSetBstMidEndSepPunct{\mcitedefaultmidpunct}
{\mcitedefaultendpunct}{\mcitedefaultseppunct}\relax
\EndOfBibitem
\bibitem[Wexler \emph{et~al.}(2014)Wexler, Heard, and Stone]{Wexler2014}
J.~S. Wexler, T.~M. Heard and H.~A. Stone, \emph{Physical Review Letters}, 2014, \textbf{112}, 1--5\relax
\mciteBstWouldAddEndPuncttrue
\mciteSetBstMidEndSepPunct{\mcitedefaultmidpunct}
{\mcitedefaultendpunct}{\mcitedefaultseppunct}\relax
\EndOfBibitem
\bibitem[Li and Cai(2014)]{Li2014}
K.~Li and S.~Cai, \emph{Soft Matter}, 2014, \textbf{10}, 8202--8209\relax
\mciteBstWouldAddEndPuncttrue
\mciteSetBstMidEndSepPunct{\mcitedefaultmidpunct}
{\mcitedefaultendpunct}{\mcitedefaultseppunct}\relax
\EndOfBibitem
\bibitem[Hui and Jagota(2014)]{Hui2014}
C.~Y. Hui and A.~Jagota, \emph{Proceedings of the Royal Society A: Mathematical, Physical and Engineering Sciences}, 2014, \textbf{470}, year\relax
\mciteBstWouldAddEndPuncttrue
\mciteSetBstMidEndSepPunct{\mcitedefaultmidpunct}
{\mcitedefaultendpunct}{\mcitedefaultseppunct}\relax
\EndOfBibitem
\bibitem[Shuttleworth(1950)]{Shuttleworth1950}
R.~Shuttleworth, \emph{Proc. Phys. Soc. A}, 1950, \textbf{63}, 444\relax
\mciteBstWouldAddEndPuncttrue
\mciteSetBstMidEndSepPunct{\mcitedefaultmidpunct}
{\mcitedefaultendpunct}{\mcitedefaultseppunct}\relax
\EndOfBibitem
\bibitem[Andreotti and Snoeijer(2016)]{Andreotti2016}
B.~Andreotti and J.~H. Snoeijer, \emph{Epl}, 2016, \textbf{113}, year\relax
\mciteBstWouldAddEndPuncttrue
\mciteSetBstMidEndSepPunct{\mcitedefaultmidpunct}
{\mcitedefaultendpunct}{\mcitedefaultseppunct}\relax
\EndOfBibitem
\bibitem[Masurel \emph{et~al.}(2019)Masurel, Roch{\'{e}}, Limat, Ionescu, and Dervaux]{Masurel2019}
R.~Masurel, M.~Roch{\'{e}}, L.~Limat, I.~Ionescu and J.~Dervaux, \emph{Physical Review Letters}, 2019, \textbf{122}, 1--6\relax
\mciteBstWouldAddEndPuncttrue
\mciteSetBstMidEndSepPunct{\mcitedefaultmidpunct}
{\mcitedefaultendpunct}{\mcitedefaultseppunct}\relax
\EndOfBibitem
\bibitem[Logg \emph{et~al.}(2012)Logg, Mardal, and Wells]{logg2012automated}
A.~Logg, K.-A. Mardal and G.~Wells, \emph{Automated solution of differential equations by the finite element method: The FEniCS book}, Springer Science \& Business Media, 2012, vol.~84\relax
\mciteBstWouldAddEndPuncttrue
\mciteSetBstMidEndSepPunct{\mcitedefaultmidpunct}
{\mcitedefaultendpunct}{\mcitedefaultseppunct}\relax
\EndOfBibitem
\end{mcitethebibliography}

\end{document}